\documentclass[aps,prb,twocolumn,showpacs,groupedaddress]{revtex4}
\usepackage{dcolumn}
\usepackage{bm}
\usepackage{graphicx}
\usepackage{natbib}
\begin{document}

\newcommand{\azu}{Cu$_3$(CO$_3$)$_2$(OH)$_2$}

\preprint{APS/123-QED}

\title{The dynamics of azurite Cu$_3$(CO$_3$)$_2$(OH)$_2$ in a magnetic field as determined by neutron scattering}
\author{K.C. Rule$^1$, D.A. Tennant$^{1,2}$, J.-S. Caux$^3$, M.C.R. Gibson$^{1,2}$, M.T.F. Telling$^4$, S. Gerischer$^1$, S. S\"ullow$^5$, M. Lang$^{6}$}
\address{
$^1$Helmholtz-Zentrum Berlin, Berlin, Germany\\
$^2$Institut f\"ur Festk\"orperphysik, TU Berlin, Berlin, Germany\\
$^3$Institute for Theoretical Physics, University of Amsterdam, Science Park 904, Postbus 94485, 1090 GL Amsterdam, The Netherlands\\
$^4$ISIS, Rutherford Appleton Laboratories, Chilton, UK \\
$^5$Institut f\"{u}r Physik der Kondensierten Materie, TU Braunschweig, Braunschweig, Germany\\
$^6$Goethe Universit\"{a}t, Frankfurt(M), SFB/TR 49, Germany\\} 

\begin{abstract}
Azurite, a natural mineral made up of CuO chains, is also an intriguing spin-1/2 quantum magnet.  There has been much debate as to the 1-dimensional (1D) nature of this material by theorists and experimentalists alike.  The focus of this debate lies in the interactions between Cu-ions within the antiferromagnetically ordered state below 1.9 K.  We present high-resolution inelastic neutron scattering data which highlight the complexity of the magnetic ground state of azurite.  The application of magnetic fields and temperatures were used to probe the excitations revealing important information about the dynamics in this system.  From this we are able to conclude that the 1D Heisenberg antiferromagnetic spin chain model is not sufficient to describe the dynamics in azurite.  Instead additional coupling including interchain interactions and an anisotropic staggered field are necessary to fully model the observed excitations.  

\end{abstract}

\maketitle

\section{Introduction}
\begin{figure}[!ht]
\begin{center}
\includegraphics[width=0.9\linewidth]{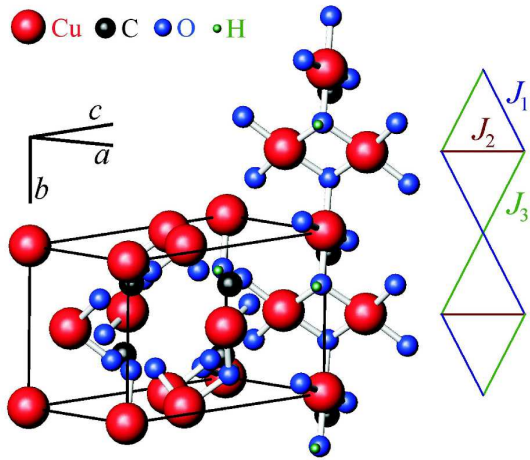}
\end{center}
\caption{(Colour online) Diamond chain model showing the structure of azurite and the relative exchange interactions within the chain.  It is now widely believed that $J_2$ is the strongest interaction in azurite.}
\label{fig:exchange}
\end{figure}

Low-dimensional magnetic systems, in particular copper oxides, are interesting subjects of study due to the novel physics that can emerge when they are cooled to low temperatures. Magnetism in these materials arises from the spin-1/2 moments carried by the Cu$^{2+}$ sites, and the magnetic couplings can vary strongly in magnitude and sign depending on the geometry and environment of the copper oxide bonds involved.  
Azurite, \azu, is one such copper oxide material which has been the subject of debate in recent years\citep{honecker, kikuchi2, gu, rule, mikeska, kang, honecker2, jeschke}.  A natural mineral, azurite can be considered one of the first experimental realizations of the 1D distorted diamond chain model.  While it orders antiferromagnetically below 1.9K\citep{spence,forstat} there has been much debate involving the relative exchange interactions between the Cu ions as shown in Fig. \ref{fig:exchange}.  Much experimental analysis and theoretical modeling have been performed however there is as yet no general consensus as to the exact magnitudes of the exchange interactions either within the 1D diamond chain, or between chains\citep{kikuchi2,gu,rule,kang,honecker}. Despite this, there is significant evidence that suggests the exchange interaction J$_{2}$ is by far the strongest interaction leading to a coupling of Cu-ions into dimers and monomers \citep{frikkee, kikuchi2, rule}.  From these results it is now widely believed that azurite cannot be described in terms of a simple isotropic exchange chain Hamiltonian, but rather that inter-chain coupling and/or anisotropic exchange must also be taken into account when describing this material.  

Recently detailed investigations of the low temperature nuclear and magnetic structures has been performed\citep{gibson, RuleNew}.  From these studies it was found that the symmetry of azurite was $P2_1$ rather than the previously determined $P2_1/c$.  For the first time, low temperature lattice parameters were found at 1.28 K to be $a = 4.99995(11)$ \AA, $b = 5.82256(14)$ \AA, $c = 10.33723(19)$ \AA\ with $\beta = 92.2103(17)^{\circ}$.\cite{RuleNew}  The magnetic moment structure was also revealed in this work where two different moments were found: $m_{0} = 0.684 \mu_{B}$ on the monomer site (Cu1) and $m_{0} = 0.264 \mu_{B}$ on the dimer site (Cu2). From this work one can now speculate how the relative magnetic moment orientation of the dimer and monomer spins can affect the local magnetic environment in the ordered phase.  For instance, it was proposed that the non-negligible moment on the dimer sites originates from the effective staggered field from the antiferromagnetically coupled monomers.\citep{RuleNew}.

In this paper, we present inelastic neutron scattering data of the quasi-one-dimensional (1D), S = 1/2 antiferromagnet (AFM) \azu.  These investigations focus on the magnetic excitations in azurite for applied fields below the plateau phase - that is below 11T when applied perpendicular to the Cu-chain direction.  We will compare our data with the ideal Heisenberg model and reveal the inconsistencies between the two.  We use the recent structural and magnetic information to devise a model which better describes azurite.


\section{Experimental method}
Inelastic neutron scattering measurements were conducted at the Helmholtz Zentrum Berlin (HZB), using FLEX, the cold-neutron triple-axis spectrometer.  A large, naturally obtained single crystal sample of \azu which has been studied previously\citep{rule} was used for all inelastic neutron scattering measurements. The sample was aligned with either an $ab$-scattering plane or a $bc$-scattering plane such that measurements could be taken along each of the principal lattice directions $H$, $K$ and $L$.  The sample was kept at 1.5 K to ensure that the system was in the ordered magnetic phase.  Constant-{\bf Q}, energy-transfer scans were performed within the range 0 - 2.5 meV with incident neutrons fixed at $k_i = 1.3$ \AA$^{-1}$. The collimation was set to guide-60'-open-open with a horizontally curved analyser. This combination of parameters gave an instrumental energy resolution of 0.11 meV. 

The high resolution time of flight spectrometer OSIRIS at ISIS, UK was also used to probe the magnetic excitations in azurite.\citep{Telling}  Magnetic fields up to 7T were applied perpendicular to the chain direction (i.e. perpendicular to the crystallographic $b^{*}$ axis).  The sample was mounted onto the end of a Kelvinox dilution insert and placed within the 7.5 T vertical field magnet.  The temperature of the sample was varied between 0.05 - 20 K but was stable for each measurement to within $\pm 0.01$ K.  The sample alignment was checked in diffraction mode, using the backscattering detector banks confirming an $a^{*}$ - $b^{*}$ scattering plane.  Data were taken using the chopper package $PG002$ which covered an energy transfer range of $-0.2 \leq$ E $\leq 2.5$ meV with a resolution at E = 0 of 0.025 meV. This energy range was chosen to maximize the resolution of the AFM spin chain which was previously observed below 2.5 meV energy transfer\citep{rule}.  
 

\section{Results}

\subsection{Dimensionality of the dynamics}
\begin{figure}[!ht]
\begin{center}
\includegraphics[width=1\linewidth]{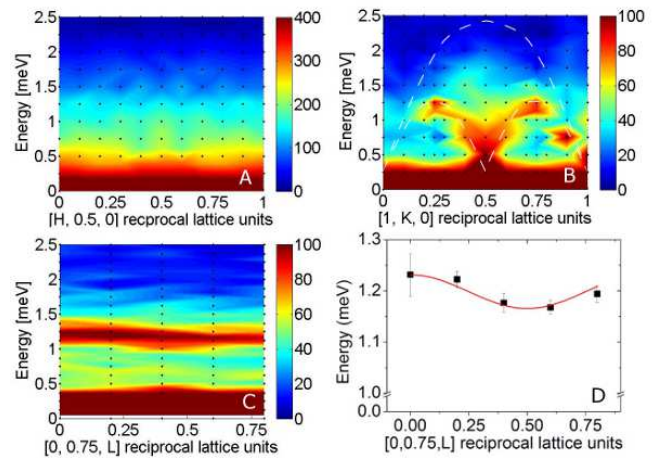}
\end{center}
\caption{Constant-\textit{Q} energy scans taken along each of the principal lattice directions $H$, $K$ and $L$ in azurite (labeled A, B and C respectively).  The intensity of each data set was normalized to the elastic intensity.  Black spots indicate the point density while smoothing has been applied.  The dashed line in B is a guide to the eye.  D shows a cosine fit to the excitation observed in C revealing an amplitude of 0.03 meV. This indicates minimal dispersion along the $L$-direction around an order of magnitude less than along the $K$-direction.}
\label{fig:FLEX}
\end{figure}

Constant-{\bf Q} energy scans were taken along each of the principal lattice directions $H$, $K$ and $L$ to ascertain the 1D nature of the low energy excitations within the ordered magnetic phase.  These measurements were taken along the [$H$, 0.5, 0], [1, $K$, 0] and [0, 0.75, $L$] directions and can be seen in Fig. \ref{fig:FLEX}. Since recent theoretical work has implied significant interchain contributions to the interactions in azurite, it was important to measure the dispersion relation along each of the three directions.  From the data in Fig. \ref{fig:FLEX}, there is little or no observed Q-dependence along the $H$ and $L$ directions, within the resolution of these measurements.  In contrast to this, there is a clear Q-dependence for measurements taken along the $K$-direction - that is parallel with the diamond chains of Cu-ions.  This is in fact the same result studied in \citep{rule} and has been attributed to the coupling of the monomer ions along the diamond chain. These results imply that, within the resolution of these measurements, the interchain couplings along the $L$ direction are at least an order of magnitude less than along $K$. 

\subsection{High resolution TOF}
\begin{figure}[!ht]
\begin{center}
\includegraphics[width=0.8\linewidth]{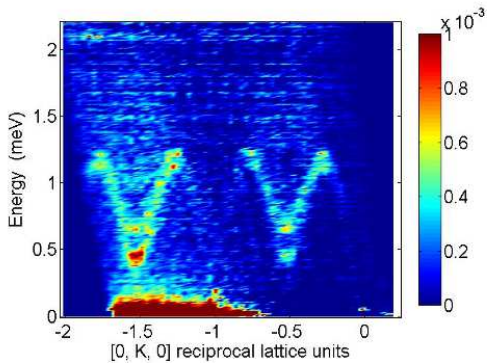}
\end{center}
\caption{Inelastic neutron scattering data covering two Brillouin zones showing the periodic dispersion along $K$.  Measurements were taken at 60 mK.}
\label{fig:WOW}
\end{figure}

The dynamics within the \textbf{HK}0 plane were measured on the single crystal sample by summing together data from 20 different sample rotation positions covering 90$^{\circ}$.  These results reveal the rich nature of the low energy magnetic scattering excitations in azurite.  It is believed that this scattering results from an effective 1D chain primarily of monomer spins, while the dimerised pairs of Cu-ions primarily contribute to scattering at higher energies.\citep{future} In this way multiple Brillouin zones could be mapped out showing the dispersive relation along the chain direction, as shown in Fig. \ref{fig:WOW}.  These data have been treated for detector efficiency with a background subtracted and the intensity modulation across the Brillouin zone is a result of the Cu-form factor. For these measurements, the sample temperature remained constant at 60 mK in order to reduce the influence of thermal fluctuations on the excitations.  While the overall dispersion looks very similar to that observed on FLEX and in earlier measurements\citep{rule}, there are clearly additional features which are revealed from the higher resolution study.  In particular, there is evidence now of 2 gaps at the AFM Brillouin zone center (0, -0.5, 0) located at $\Delta_1 \approx 0.4$ meV and $\Delta_2 \approx 0.6$ meV. While the periodic dispersion resembles the scattering from a 1D Heisenberg Antiferromagnetic chain (HAFC) with a spinon continuum, this model is gapless as demonstrated in the simulation of Fig. 4c.  Heisenberg interchain interactions would allow for the observed gaps in the $K$-dispersion however these would also be responsible for dispersive behaviour along the other principal directions, $H$ and $L$.  Since Fig. \ref{fig:FLEX} showed almost no dispersion, then we must consider an anisotropic interchain interaction which would not be dispersive perpendicular to the chain.  

\subsection{Temperature variation of the low energy band}
\begin{figure}[!ht]
\begin{center}
\includegraphics[width=1.0\linewidth]{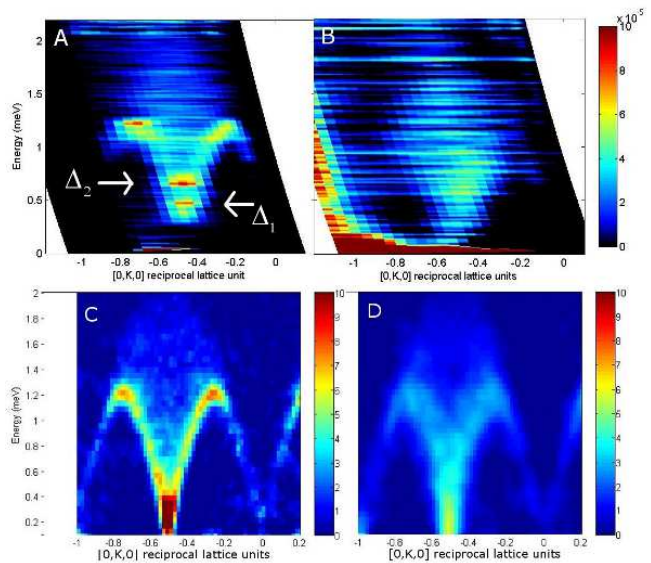}
\end{center}
\caption{The dynamic scattering from azurite along the Cu-chain direction in azurite is shown at 0.04 K (A) and 2.5 K (B).  The periodic dispersion in $K$ at 40 mK resembles the 1D HAFC with a spinon continuum however this model does not allow for the energy gaps, $\Delta_1$ and $\Delta_2$, clearly observed in this high resolution data (indicated by the arrows). The lower plots show the expected scattering from a 1D HAFC at 0.04 K (C) and 2.5 K (D).  The absence of gaps from the model in C indicates that interchain coupling and anisotropy terms play a significant role in the azurite Hamiltonian. In contrast the data and simulation taken above the magnetic ordering temperature look quite similar. }
\label{fig:Tdep}
\end{figure}

Focusing on one Brillouin zone with higher statistics, the temperature dependence of the low energy scattering can be observed, as seen in Fig. \ref{fig:Tdep}. The two gaps at the zone center are clearly evident in the lowest temperature data set, however at higher temperatures they are not observed.  At 2.5K azurite is no longer in its 3D magnetically ordered phase.  However scattering from the diamond chain and the spinon continuum is still evident. In particular the scattering is strongest at the zone center with a possible sinusoidal dependence resembling that of the 1DHAFC model.  It is clear that while the signal is much weaker at 2.5K, the chain dynamics still persist.  While the counting statistics of the higher temperature data sets are insufficient for a quantitative assessment, they show a qualitative agreement with the 1D spin chain model.  

\subsection{Applied field TOF}
\begin{figure}[!ht]
\begin{center}
\includegraphics[width=1.0\linewidth]{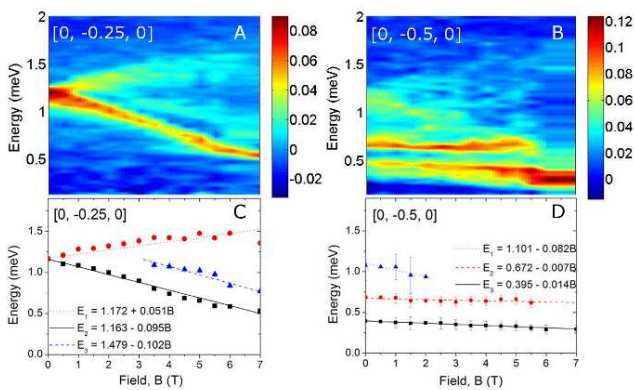}
\end{center}
\caption{The field dependence of the low energy excitations taken at the zone boundary (A) with $Q = [0,-0.25,0]$ and the zone centre (B) with $Q = [0,-0.5,0]$.  Gaussian fits to the excitations show a divergence of the modes at the zone center (C) while each of the three modes at the zone boundary (D) show different field dependent behaviour.  These data indicate that the nature of the excitations is not that of standard spinwaves in an isotropic antiferromagnet.}
\label{fig:Bdep}
\end{figure}

By applying a magnetic field we can influence the magnetic dynamics in such a way as to reveal more about the interactions within azurite. The dynamics of azurite were observed as a function of magnetic field up to 7T applied parallel to the $c$-direction.  The zone center and boundaries were observed to show the most notable changes and have been plotted as a function of field at 0.04K in Fig. \ref{fig:Bdep}. The zone-boundary mode at a zero field energy of 1.2meV clearly splits into 2 modes at low fields with a third mode splitting from around 3T.  While these modes diverge with increasing field, they do not follow a Zeeman type splitting.  


At the zone-center, three modes are clearly visible with varied behaviour in field.  The weak upper mode at $\approx 1.2$ meV tracks the zone boundary behaviour of Fig \ref{fig:Bdep} A and could be a shadow due to secondary scattering processes from incoherent scattering.   The lower gap, $\Delta_1$, shows a clear decrease in energy with applied field while the upper gap, $\Delta_2$, remains approximately constant with field. The different field behaviors indicate that the origin of the two gaps also differs as will be discussed later.

\begin{figure}[!ht]
\begin{center}
\includegraphics[width=1.0\linewidth]{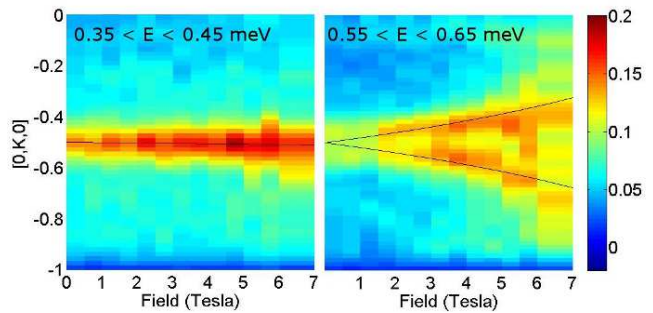}
\end{center}
\caption{The variation in $K$-vector of the zone-center energy gaps with applied field.  Left: the lower gap located in the range $0.35 < En < 0.45$ meV shows almost no change in $K$ with field (as indicated by the black line) until 6T when Gaussian fits indicate a small splitting. Right: the upper gap in the range $0.55 < En < 0.65$ meV shows a more significant splitting at low fields. The black lines follow the magnetization per spin of the Heisenberg chain as defined by equation (1). }
\label{fig:Bdep2}
\end{figure}

The effect of the applied field on the Q-vector of the energy gaps can reveal information about the character of the two different modes. Fig. \ref{fig:Bdep2} shows how the lattice vector of the dynamics behaves in fields up to 7 T.  The lower gap, which has been displayed by binning the energy range $0.35 < En < 0.45$ meV, shows almost no deviation from the zero field value.  
The gap at higher energies $0.55 < En < 0.65$ meV shows a different field dependence with a clear splitting of $K$ from around 1 T. As discussed below, this suggests that the lower mode has originated more from the transverse continuum (spin fluctuations perpendicular to the field) and the upper from the longitudinal continuum (spin fluctuations along the field).  

\section{Analysis and Discussion}

\subsection{Comparison to the 1D HAFC model in a field}
 
\begin{figure}[!ht]
\begin{center}
\includegraphics[width=0.8\linewidth]{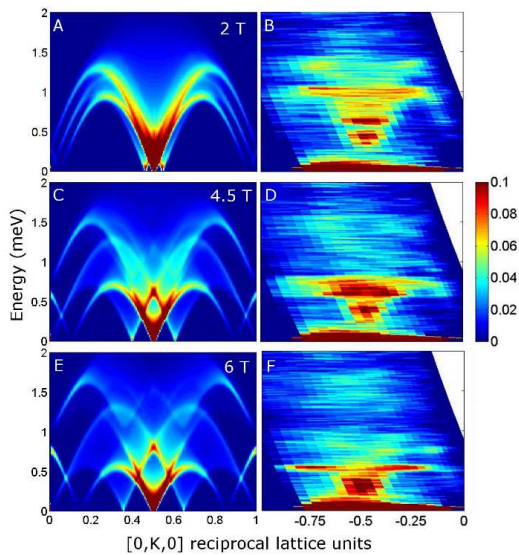}
\end{center}
\caption{A, C and E show simulations of the inelastic neutron scattering data for applied fields 2T, 4.5T and 6T.  These can be compared to the experimental results for the same fields in B, D and F. A background from the zero field data set has been subtracted from the data on the right.}
\label{fig:Bdepsimexp}
\end{figure}

To gain more insight into the physics of azurite we first consider the effect of applied field and how this influences the spin 1/2 Heisenberg chain model.  In Fig. \ref{fig:Bdepsimexp} A, C, and E we demonstrate how applied field will affect the spin correlations for the 1D Heisenberg model.  These simulations involve taking the average contribution from the spin correlations along the field direction $S^{||}(Q,\omega)$ and perpendicular $S^{\perp}(Q,\omega)$.  The spin correlations are computed for finite chains of 192 sites using the integrability-based ABACUS method \cite{2005_Caux_PRL_95,2005_Caux_JSTAT_P09003,2009_Caux_JMP_50}. 
Overall, the general features observed in $S(Q,\omega)$ compare favorably with the neutron scattering data shown in Fig. \ref{fig:Bdepsimexp} B, D, and F despite the absence of gaps. However it is clear that the rotational symmetry in azurite is broken by the applied field. 

\begin{figure}[!ht]
\begin{center}
\includegraphics[width=0.8\linewidth]{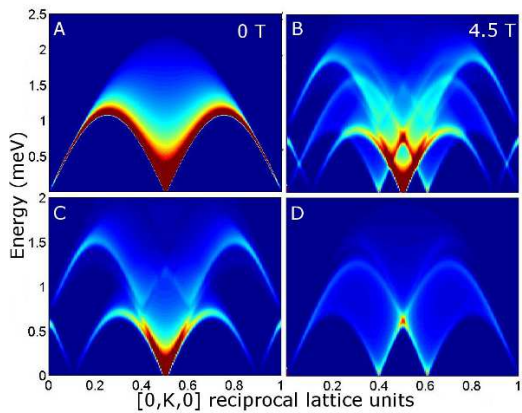}
\end{center}
\caption{Simulation of the 1D HAFC at zero field (A) and at 4.5T (B).  The 4.5 T simulation can be separated into the two components: transverse (C), indicating the spin correlations perpendicular $S^{\perp}(Q,\omega)$, and longitudinal (D), indicating the spin correlations along the field direction $S^{||}(Q,\omega)$.}
\label{fig:Simul}
\end{figure}

In zero field there is only one contribution to the scattering $S(Q,\omega)$ for the Heisenberg model which is shown in Fig. \ref{fig:Simul} A.  However in an applied field, the separate contributions from $S^{||}(Q,\omega)$ and $S^{\perp}(Q,\omega)$ combine to give the scattering pattern in Fig. \ref{fig:Simul} B for 4.5 T.  Here, we define $S^{||}(Q,\omega)$ as the spin correlations longitudinal to the applied field and $S^{\perp}(Q,\omega)$ as the average of the spin correlations transverse to the applied field.   In contrast to zero field, the 4.5 T scattering from the transverse and longitudinal correlations can be considered separately as in Fig. \ref{fig:Simul} C and D respectively.  

\begin{figure}[!ht]
\begin{center}
\includegraphics[width=0.8\linewidth]{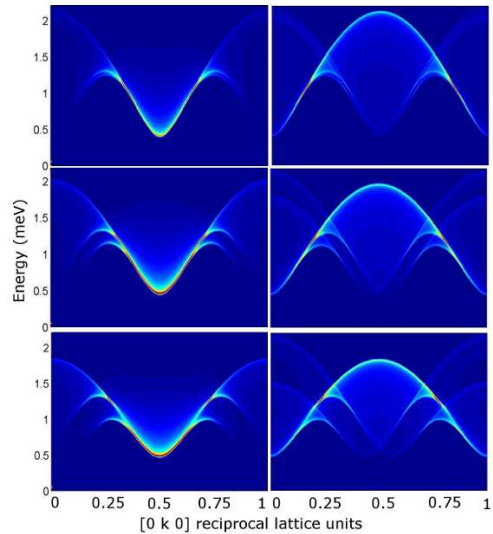}
\end{center}
\caption{Simulation of the anisotropic Heisenberg Hamiltonian, as defined in Equation 1 from Ref. \cite{hagemans}, where $J_x = 1$, $J_y = 0.5$ and $J_z = 1$.  The transverse (left column) and longitudinal (right column) continua scattering have been plotted for applied magnetic fields (from top to bottom) $\textbf{H}_z=0.01, 0.05, 0.1$.  While the zone centre of the transverse mode shows no incommensuration along $K$, the longitudinal mode shows significant splitting.  Both show a strong field dependence at the zone boundary.}
\label{fig:BTL}
\end{figure}

The transverse and longitudinal continua behave very differently in applied fields and have been modeled in Fig. \ref{fig:BTL}.  Here the transverse (left panels) and longitudinal (right panels) continua have been calculated for an anisotropic Heisenberg Hamiltonian, as defined in Equation 1 from Ref. \cite{hagemans}, with applied fields increasing up to $\textbf{H}_z = 0.1$. Here the saturation field along $\textbf{H}_z$ (equivalent to 11 T for azurite) is $J_x$, $J_y$ and $J_z$ dependent.  The anisotropy in this XYZ model is responsible of the energy gap observed at the zone centre. As can be seen, the transverse mode does not split in $K$ at the zone centre whereas the longitudinal does. However both show significant splitting at the zone boundary. Since the lower mode observed in azurite (0.4meV) (Fig. \ref{fig:Bdep2}) does not appear to split in $K$ with field we can deduce that this originates from the transverse continuum while the upper gapped mode (0.6meV), which does split with applied field, originates from the longitudinal continuum. 

\begin{figure}[!ht]
\begin{center}
\includegraphics[width=0.8\linewidth]{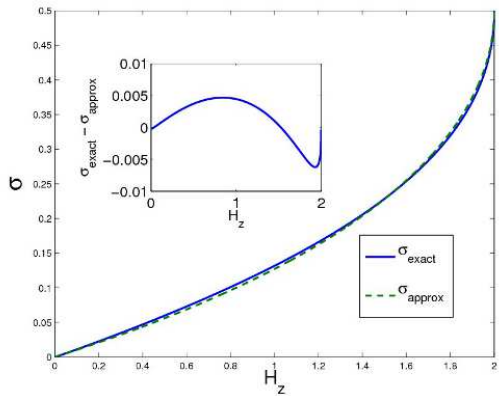}
\end{center}
\caption{(color online) Plot of the magnetization curve for the isotropic XXX antiferromagnet. The blue line is the exact curve obtained from the Bethe Ansatz. The green dashed curve is the approximation in Eq. 1. The subplot gives the difference between exact and approximate curves, which remains small throughout the parameter range considered.}
\label{fig:caux}
\end{figure}

The magnetization per spin in a spin-1/2 Heisenberg chain versus field can be approximated using the following equation from Ref.\cite{Muller}: 
\begin{eqnarray}
\sigma(h) = \frac{1}{\pi} a\sin\frac{1}{1-\pi/2 + \pi h_{sat}/2h} 
\end{eqnarray}
where $0 < h < h_{sat}$ and $h_{sat} = 11$ T for azurite. The incommensuration along $K$ (to be compared to the right panels of Fig. \ref{fig:BTL} for the longitudinal mode) is $q(h) = 1/2 \pm \sigma(h)$. 
This uses the relationship between magnetization and filling factor for spinons. It also includes the field dependent quantum renormalization factors. The incommensuration should go from 0.5 in zero field to 0.35 and 0.65 by 7 Tesla in azurite and be approximately linear in this field region. This looks similar to what is actually observed in azurite in Fig. \ref{fig:Bdep2} and confirms that the approximation of Eq. 1 is sufficiently accurate with respect to the exact solution as seen in Fig. \ref{fig:caux}.

\begin{figure}[!ht]
\begin{center}
\includegraphics[width=0.8\linewidth]{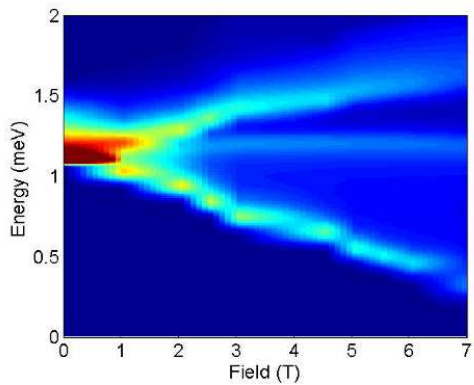}
\end{center}
\caption{Simulation of the inelastic neutron scattering data from Fig. \ref{fig:Bdep} A at the zone boundary in applied fields up to 7T.}
\label{fig:BdepZB}
\end{figure}

To see more clearly how the continuum boundaries shift with applied field, a simulation of the inelastic neutron scattering data is plotted in Fig. \ref{fig:BdepZB} showing the field dependence of all modes together and taken as a cut along $K$ between 0.22 and 0.28. This calculation captures important features of the modes that were seen in Fig. \ref{fig:Bdep}. From the mode splitting at the zone boundary one can deduce that the character of the observed excitations is not that of standard spinwaves in an isotropic antiferromagnet. In spinwave theory the excitations are oscillatory modes which are not changed significantly at the zone boundary. The strong zone boundary behaviour with field is often seen for spinon systems with strong quantum fluctuations\cite{coldea3}.

It is clear from the current high-resolution results that the simple 1D Heisenberg antiferromagnetic chain model is insufficient to fully describe azurite. When determining a suitable model to describe the excitation spectra, additional terms such as anisotropy and interchain coupling must be considered.   Despite the broad similarities between the 1D HAFC model and the azurite data, these additional terms are more effective at describing the details in the neutron scattering data.


\subsection{Anisotropy}
As observed in Fig. \ref{fig:BTL}, by introducing anisotropy into the system, an energy gap opens up in the excitation spectra.  This can be modeled by the anisotropic XYZ-Hamiltonian which can be written as:

\begin{eqnarray}
H = \displaystyle\sum\limits_{i}J_zS^z_iS^z_{i+1} + \frac{(J_x + J_y)}{4} (S^+_i S^-_{i+1} + S^-_i S^+_{i+1}) \nonumber \\
 + \frac{(J_x - J_y)}{4} (S^+_i S^+_{i+1} + S^-_i S^-_{i+1})	
\end{eqnarray}

where $S^z_i$ and $S^z_{i+1}$ are the spin matrices at neighbouring sites and the exchange interactions $J_x \neq J_y \neq J_z$.  Here, the excitations in the pure XYZ chain are gapped spinons which form an empty particle and filled hole band \citep{caux}. Neutrons measure continua of pairs of these excitations just as in the isotropic case. Although the spherical rotational symmetry is completely broken there is only one type of spinon which is now gapped. Within a mean field theory the particle and hole dispersions are given by \citep{hagemans}:

\begin{equation}
\omega (k) = \pm \sqrt{(t\cos{(k)}-h)^2 + \Delta^2\sin^2{(k)}} 
\end{equation}

where $t$ and $\Delta$ are calculable from the $J_x$, $J_y$ and $J_z$ and $h$ is the magnetic field. The transverse continuum of scattering is given by pair-states of momenta $k_1$ and $k_2$ at $q = k_1 + k_2 + \pi$ and energy $\omega = \omega(k_1) + \omega(k_2)$. Whereas the longitudinal continuum moment is $q = k_1 - k_2$ with the same energy relation $\omega = \omega(k_1) + \omega(k_2)$. The energy gap for zero field is the same for the longitudinal and transverse continuum so does not explain the different gaps in the continua that we actually observed.

\subsection{Interchain coupling}

While anisotropy is likely to be responsible for opening one energy gap in the ground state excitation spectrum, interchain coupling can be used to describe the multiple gaps observed.  The effects of interchain coupling in azurite have previously been down played due to the minimal dispersion observed in the earlier inelastic neutron scattering results\citep{rule}, however more recently the importance of interchain coupling has been highlighted as necessary to fully describe the interactions between Cu-ions.\citep{jeschke, kang} In fact, some form of anisotropic interchain coupling may also account for the slight dispersion between the chains, observed in Fig. \ref{fig:FLEX}.
For KCuF$_3$, a model HAFC with weak interchain coupling, the quantum field theory works almost perfectly and is accurate to a few percent. In contrast, for azurite the agreement with the quantum field theory for coupled Heisenberg chains is only within ~30\%. This means that an anisotropic interchain coupling (Ising like) is a credible possibility and interchain couplings should be considered to have a similar influence to within-chain anisotropy.

Interchain coupling theories which have been developed for the Heisenberg case, may also be applied for the case of Heisenberg chains coupled through an Ising interchain interaction. These theories can give a reasonable approximation of interchain coupling involved in azurite. The formula can be adapted to explain the N\'eel ordering temperature $T_{N} = 1.9$ K, ordered moment $\langle S \rangle = 0.684 \mu_{B}$ 
and gap energy $\Delta =0.40$ meV and can all be related to the interchain coupling $J_{\perp}$ by using quantum field theory. 
For instance, we can estimate the interchain coupling by considering the relationship between the energy gap and the ordered monomer moment as outlined in Essler et al., \citep{essler}.  A combination of equations 15 and 56 from this paper gives the simple relation $|J_{\perp}| = 1/J (0.175[\Delta/m_{0}])^2 = 0.048$ meV.  Similarly, taking the approach of Schulz\citep{schulz}, the interchain interaction can be approximated as 0.065 meV for a gap $\Delta = 0.4$ meV. If we consider the critical temperature, $T_{N} = 1.9$ K, to influence the strength of the interchain coupling, we find that for a cubic lattice the interchain coupling would be $|J_{\perp}| = 0.083$ meV as outlined in the work of Bocquet\citep{bocquet}.  A more sophisticated analysis by Irkhin and  Katanin\citep{irkhin} gives a slightly improved estimate of the interchain coupling as $|J_{\perp}| = 0.086$ meV with the following formula $m_{0} \approx 1.085(J_{\perp}/J)^{1/2}$.

The effect of three dimensional ordering of an array of spin chains is to induce an effective staggered field on the chains.  In azurite, the spinons in the continuum are bound into well defined modes by the interchain coupling.  Within this spinon picture for the Heisenberg chain, the staggered field produces two modes, one with gap $\Delta_1$ which is a doublet which has a polarization perpendicular (transverse) to the staggered field (ordering direction). A breather mode (consisting of 2 bound solitons, or 4 bound spinons) also forms with a gap $\Delta_2$ which has a polarization along the staggered field direction (longitudinal).  In the special case of pure Heisenberg interaction the gap ratio $\Delta_1$:$\Delta_2$ is 1:$\sqrt{3}$. In general this ratio will vary with anisotropy for an XYZ Hamiltonian. For the case where the applied field and staggered field directions coincide then the lower mode with gap $\Delta_1$ which has a transverse polarization will remain commensurate as is observed, whilst the upper mode with $\Delta_2$ will split in $Q$ with field as also observed.  

In the current experiments the applied field was along the $c$-axis when the magnetic ordering was observed for the monomers to be in the $ac$ plane at an angle of $55^\circ $ from $c$\citep{RuleNew}. 
The spin fluctuations along the $b$-axis then are both transverse to the field and to the magnetic order. They then should not split in $q$ and will have a lower gap energy that changes only weakly with field. Neutron measurements are sensitive to the fluctuations perpendicular to $Q$ and since the chain direction ($b$) was aligned almost completely perpendicular to the wavevector transfer, the $b$-axis mode (transverse) is more strongly probed rather than the $a$ correlation. The correlations for the $c$ component (longitudinal) will always appear strongly because they are vertical and therefore always perpendicular to $q$ transfer. These will pick up the incommensuration from the longitudinal nature of the field in relation to the ordering and so we expect incommensuration of the upper mode as is seen. 

In the spin chain material CDC, which is a fundamentally Heisenberg system with very small interchain coupling \citep{kenzelmann}, similar behaviour to azurite is seen. In particular solitons and breathers are observed as well as the incommensurate behaviour of these modes. However the situation is not identical to azurite which is probably an XYZ chain with larger interchain coupling.  Also, in CDC the staggered field is due to the applied field combining with the Dzyaloshinskii-Moriya interaction whereas for azurite the staggered field is probably from an interchain coupling effect.

Finally, the most powerful method to determine the exchanges in a quantum magnet is to use the method of saturation (which was developed to understand Cs$_2$CuCl$_4$ in \citep{coldea2}). So far only the exchange couplings along the chain direction in azurite have been observed by applying fields above 11 Tesla\cite{rule}. It is interesting to note that the energy gap at the zone center does not close at the onset of the plateau phase ($H_{c1} \approx 11$ T). This is a signature of anisotropy and/or interchain coupling. The scale of this gap at 11 T was around 0.142 meV giving an interchain coupling of $J_{\perp} \approx 0.071$ meV which agrees well with the values deduced above.  In addition the scattering at the zone center, $q =(1,0,0)$, when linearly extrapolated as a function of field (as seen in Fig. 4 from \citep{rule}) also does not close at 0 T. A possible explanation for this involves interchain coupling from the (1,0,0) that is ferromagnetic along 110, 100, and 1-10 directions. This requires more detailed analysis but may indicate the presence of frustrated interchain interactions. To understand this saturation phase better, the dispersions along the $a$ and $c$ directions should be measured such that the couplings within the plateau phase can be extracted.


\section{Conclusion}

In summary, high resolution inelastic neutron scattering results have shown that azurite is not well described by the 1D HAFC model.  Instead, more complex interaction pathways are a necessary addition to the spin-chain Hamiltonian to describe the observed features.  Anisotropic interchain interactions of around 10\% of the intrachain interactions are required to explain the two gapped modes in the dispersion along the chain direction.  The 3D long range magnetic ordered state can also be attributed to the presence of these interchain interactions.  

\begin{acknowledgments}
We would like to thank A. Honecker, and D. Le for useful discussions.  The authors are grateful for the local support staff at ISIS, HZB and the ILL. This research was supported in part by the Deutsche Forschungsgemeinschaft DFG under Grant No. SU229/9-1. JSC acknowledges support from the Foundation for Fundamental Research on Matter (FOM), which is part of the Netherlands Organisation for Scientific Research (NWO).
\end{acknowledgments}

\end{document}